\documentclass[twoside,twocolumn,english,prl,aps,superscriptaddress,showpacs]{revtex4}
\usepackage[T1]{fontenc}
\usepackage[latin9]{inputenc}
\usepackage{color}
\usepackage{babel}
\usepackage{textcomp}
\usepackage{amsmath}
\usepackage{amssymb,amsfonts,mathrsfs}
\usepackage{amsthm}
\usepackage{graphicx}
\usepackage[unicode=true,pdfusetitle,
 bookmarks=true,bookmarksnumbered=false,bookmarksopen=false,
 breaklinks=false,pdfborder={0 0 0},backref=false,colorlinks=true]
 {hyperref}
\usepackage{breakurl}
\usepackage{epstopdf}

\makeatother
\theoremstyle{plain}

\makeatother

\begin{document}

\preprint{This line only printed with preprint option}

\title{Link between \emph{Zitterbewegung} and topological phase transition}

\author{Xin Shen}
\email{shenx@cjlu.edu.cn}
\affiliation{College of Sciences, China Jiliang University, Hangzhou 310018, China}

\author{Yan-Qing Zhu}
\email{yqzhuphy@hku.hk}
\affiliation{Department of Physics and Center of Theoretical and Computational Physics, The University of Hong Kong, Pokfulam Road, Hong Kong, China}
\affiliation{Guangdong-Hong Kong Joint Laboratory of Quantum Matter, Frontier Research Institute for Physics, South China Normal University, Guangzhou 510006, China}

\author{Zhi Li}
\email{lizphys@m.scnu.edu.cn}
\affiliation{Guangdong-Hong Kong Joint Laboratory of Quantum Matter, Frontier Research Institute for Physics, South China Normal University, Guangzhou 510006, China}
\affiliation{Guangdong Provincial Key Laboratory of Nuclear Science, Institute of Quantum Matter, South China Normal University, Guangzhou 510006, China}
\affiliation{Guangdong Provincial Key Laboratory of Quantum Engineering and Quantum Materials, SPTE, South China Normal University, Guangzhou 510006, China}

\begin{abstract}
Topological quantum state described by the global invariant has been extensively studied in theory and experiment. In this letter, we investigate the relationship between \emph{Zitterbewegung} and the topology of systems that reflect the properties of the local and whole energy bands, respectively. We generalize the usual two-band effective Hamiltonian to characterize the topological phase transition of the spin-$J$ topological insulator. By studying \emph{Zitterbewegung} dynamics before and after topological phase transition, we find that the direction of quasiparticles' oscillation can well reflect topological properties. Furthermore, we develop a quantitative calculation formula for the topological invariant in the spin-$J$ Chern insulator and give the selection rule of the corresponding dynamics. Finally, we demonstrate that our theory is valid in different topological systems. The topological invariant can be represented by local dynamical properties of the high-symmetry points in the first Brillouin zone, which provides a new measurement method from the dynamical perspective.
\end{abstract}


\maketitle

\emph{Introduction.}---As a state of matter beyond the conventional symmetry breaking paradigm in terms of the classification of phases, topological quantum state has long been a hot topic in condensed matter physics~\cite{topo_zoo}. Topological materials, due to their robust edge mode, boast very good fault tolerance as quantum computing devices~\cite{hasan_ti,xlqi_ti,topo_computation}. In the past decade, topological insulators, topological superconductors and topological semi-metals have been experimentally realized one after another~\cite{hasan_ti,xlqi_ti,topo_material}. In addition to solid materials, quantum simulation has been currently applied in some relatively pure and controllable artificial systems inclusive of photonic crystal~\cite{topo_photonics}, superconducting qubits~\cite{Roushan2014,topo_qubit} and ultracold atomic gases~\cite{haldanecold,topo_coldatom1,topo_coldatom2} etc., to facilitate further study on topological quantum state.

Since the topology of a system is characterized by the non-local invariant, it is essential in experiment to detect the topological invariant. In condensed matter physics, the mainstream scheme for obtaining the topological invariant is indirect~\cite{topo_order}, because the measurement of the wave function per se is far from easy~\cite{Hsieh2008,Xia2009,hasan_ti,xlqi_ti}. In artificial quantum systems (such as ultracold atoms, trapped ion, photonic crystals etc.), however, multiple methods can be developed due to their high controllability. Measurement method for the topological invariant varies in different dimensions. In one-dimension (1D), the topological invariant or Zak phase can be measured through Ramsey interference~\cite{zak_phase}. While in 2D, the topological invariant can be experimentally obtained by band tomography (such as Wilson lines~\cite{wilsonlines} and quench dynamics~\cite{tomography1}), because Berry curvature works well as a magnetic field causing the transverse drift of quasiparticles, and the Berry curvature itself is determined by the eigenstate wave function of the corresponding band. Recent reports have it that the system can be quenched into different topological states through dynamical means so as to obtain the topological properties of the system~\cite{quench1,quench2,quench3}.

In this letter, we uncover the link between \emph{Zitterbewegung} (ZB)~\cite{schrozb} and topological phase transition, and propose a dynamical scheme to detect the topological invariant via the quasiparticles' motion behavior. Generally it is demonstrated that the mechanism behind ZB phenomenon is the inter-band interference, which is ubiquitous in multi-band systems~\cite{generalZB}. Two-band and multi-band topological insulators are no exception and the topological phase transition in the Chern insulator is usually characterized by the process of band closing-and-reopening. 

To reveal the ZB dynamics' change during the band inversion process, we first generalize the two-band model, which can be used to characterize the band inversion of the Chern insulator, to an arbitrary spin-$J$ system that depicts spin-$1$ Maxwell quasiparticles~\cite{yqzhu,yqzhu2}, spin-$3/2$ Rarita-Schwinger-Weyl quasiparticles~\cite{spin321,spin322,spin323} or even multi-band systems with higher spin. Meanwhile, we find the selection rule for spin-$J$ system's ZB effects, i.e., only adjacent bands can induce ZB. In addition, due to the fact that the global topological invariant defined in the first Brillouin zone can always be characterized by local topological indices of high-symmetry points (HSPs)~\cite{xjLiu}, together with the one-to-one correspondence between ZB dynamics and local topological indices near HSPs proved in this letter, we reach the final conclusion that the whole topological characteristics can be reflected by ZB dynamics. Therefore, it is universally applicable to obtain the topological invariant through ZB dynamics for topological materials of different symmetric classes~\cite{Chiu2016}. 	

As an illustration and  application, we further discuss the concrete ZB dynamics in the well-known 2D Kane-Mele model~\cite{z2kanemele} and 3D chiral topological insulators~\cite{3dchiralti}. With the advantage of our scheme, we can obtain the topological properties of the whole system by simply calculating dynamical behaviors of the quasiparticle near the HSPs. Note that, the amplitude of ZB is inversely proportional to the width of the energy gap, therefore, the closer the parameters are to the phase transition point, the easier the ZB effect is to be observed, and thus the more accurate the measurement results are~\cite{quench3}.

\emph{General Theory.}---ZB originates from the interference between different bands, whereas the topological phase transition is always marked by the band inversion. The two seemingly ``chalk-and-cheese'' physical mechanisms, ZB and topological phase transition, are actually intriguingly linked with each other for both are bound up with the energy band formula. Hamiltonian of an arbitrary multi-band system can be written as $H(\mathbf p)$, while the corresponding position of center of mass (PCM, which suggests ZB phenomenon) takes the form~\cite{generalZB}
\begin{equation}
\label{rot1}
\mathbf{r}(t)=\mathbf{r}(0)+t\sum_m\mathbf{V}_mQ_m+i\sum_{m\neq n}e^{i\omega_{mn}t} \frac{   Q_m \frac{\partial H}{\partial \mathbf{p}}  Q_n }{E_n-E_m} ,
\end{equation}
where the Hamiltonian is diagonalized as $H=\sum E_m Q_m$ with the projection operators  $Q$. The second term is the usual velocity operator where $\mathbf{V}_m=\frac{\partial E_m}{\partial \mathbf{p}}$ and in the last term the ZB frequencies are $\omega_{mn}=E_m-E_n$. The projection operators $Q_m$, $Q_n$ ($m\neq n$) in the last term indicate that ZB comes from the band interference. From the expression, we notice that the oscillatory term can be regarded as an aggregate of all pairs of bands. Therefore, the multi-band model can be reduced to a two-band one without loss of generality. The generalization to the multi-band model is straightforward by including all the possible two-band combinations.

For simplicity, we consider only the ZB term of a two-band model and rewrite the expression as
\begin{equation}
\mathbf r_o(t)=\frac{1}{\omega}\left(-ie^{i\omega t}Q_a\frac{\partial H}{\partial \mathbf p}Q_b+ie^{-i\omega t}Q_b\frac{\partial H}{\partial \mathbf p}Q_a \right),
\end{equation}
where $\omega\equiv E_a-E_b$. Due to the high frequency and small amplitude, ZB phenomenon of elementary particles is difficult to observe experimentally. However, it can be studied in artificial systems (e.g., trapped ion, optical crystal and ultracold gases) by means of quantum simulation. For a given initial state $|\psi_\mathbf p(0)\rangle$, the corresponding trajectory of PCM reads
\begin{equation}
\label{rot}
\langle \mathbf r_o(t) \rangle=\frac {2\mathcal A}{\omega} \cos(\omega t+\theta),
\end{equation}
where $\langle -iQ_a\frac{\partial H}{\partial \mathbf p}Q_b\rangle\equiv \mathcal Ae^{i\theta}$. Note that, both $\mathcal A=(\mathcal A_x,\mathcal A_y,\mathcal A_z)$ and $\theta=(\theta_x,\theta_y,\theta_z)$ in the above expression are vectors. The phase transition in topological insulators usually concurs with the process of band closing-and-reopening, which finally leads to band inversion near the degenerate point. As to the expression of ZB, the projection operators $Q_a$ and $Q_b$ stay perfectly intact during band inversion, while the energy gap parameter $\omega$ changes in not only the absolute value but also its sign. After band inversion, trajectory of the quasiparticles with the same initial state can be expressed as
\begin{equation}
  \langle \mathbf r_o(t) \rangle=-\frac {2\mathcal A}{\omega} \cos(\omega t-\theta).
\end{equation}
In the above equation, the trajectory echoes the phase angle $\theta$. When $\theta=0$, the quasiparticle's PCM will move in opposite directions before and after the phase transition, marking an significant change that can be used to characterize the topological phase transition. Note that, we can always extract the sign of the energy gap parameter $\omega$ for arbitrary value of $\cos\theta$, except $\theta=\pi/2$. In the special case of $\theta=\pi/2$,  the quasiparticle's PCM  can still reflect the topological phase transition. Next, we will use spin-$J$ model to explain in detail the generality of our theory.

\emph{Spin-$J$ System.}---Under low-energy approximation, the 2D Chern insulator can be described by the two-level effective Dirac Hamiltonian. Here we generalize the 2D Chern insulator case to arbitrary spin-$J$ case, i.e.,
\begin{equation}
\label{hamJ}
H=v_xp_xJ_x+v_yp_yJ_y+mJ_z,
\end{equation}
where $\mathbf J$ is the angular momentum operator, which satisfies $[J_i,J_j]=i\epsilon_{ijk}J_k$, and $m$ is the mass term proportional to the width of the energy gap. As the $m$ term changes from negative to positive, the band inversion is taking place near $\mathbf p=0$, and the system is undergoing topological phase transition in the meantime. Before we proceed analytically, it is necessary to emphasize that the model is general. In addition to the 2D Chern insulator, the model can also be applied to the 1D and 3D spin-$J$ systems. For example, when we take $v_x=0$ or $v_y=0$, it describes a Su-Schrieffer-Heeger model~\cite{zak_phase} and when $m$ is replaced by $v_z p_z$, it describes the behavior of Weyl semi-metal~\cite{weylfermion,Xu613}.

In general, when we take the initial state $|\psi_\mathbf p(0)\rangle$ with finite momentum $\mathbf p$, ZB effect of the quasiparticle will decay very fast. Experimentally, the state located at the avoided crossing point is more suitable to be chosen as the initial state, which has zero group velocity. Therefore, we consider the initial state of $\mathbf p=0$, which is coincident with the locus of band inversion, and obtain the trajectory of ZB caused by pairs of adjacent energy bands, i.e.,
 \begin{equation}
 \begin{split}
 \langle x_o(t)\rangle=&\frac{\mathcal A_x}{m} \cos(mt+\theta)\\
 \langle y_o(t)\rangle=&\frac{\mathcal A_y}{m} \sin(mt+\theta),
 \end{split}
 \end{equation}
where
 \begin{equation}
 \begin{split}
 \mathcal A_{x,y}e^{i\theta}=&\langle \psi(0)|a\rangle\langle b|\psi(0) \rangle \\ &\times -iv_{x,y}  \sqrt{(J+1)(a+b-1)-ab}
 \end{split}
 \end{equation}
with $|b - a|=1$. From the above equation, we find that the trajectory is actually like a circle drawn counterclockwise on the $xy$-plane. When band inversion occurs, i.e., $m\rightarrow -m$, the trajectory expression turns into
 \begin{equation}
 \begin{split}
 \langle x_o(t)\rangle=&-\frac{\mathcal A_x}{m} \cos(mt-\theta)\\
 \langle y_o(t)\rangle=&\ \ \ \ \frac{\mathcal A_y}{m} \sin(mt-\theta),
 \end{split}
 \end{equation}
which also circles a loop but in an opposite direction to that before band inversion. This qualitative behavior is $\theta$ independent, which allows ZB to well characterize topological phase transition. For the spin-$J$ system, unlike the usual multi-frequency ZB~\cite{generalZB}, we prove that only these adjacent bands will induce ZB effect. Let's name it "\emph{Selection rule of ZB}", whose mathematical proof is shown in supplementary material S1.
By setting the linear combination of eigenstates as the initial state, one can map the inversion process. To be more specific, for the 2D Chern insulator, different phases of the system can be characterized by the topological Chern invariant. It has been proved that this invariant can be extracted from the ``topological charge'' of the Dirac point~\cite{Ch12}. Simultaneously, the band inversion near the Dirac point will reverse the direction of ZB, which establishes the connection between the topological invariant and ZB dynamics.

\begin{figure}[htbp]
\begin{center}
\includegraphics[scale=0.45]{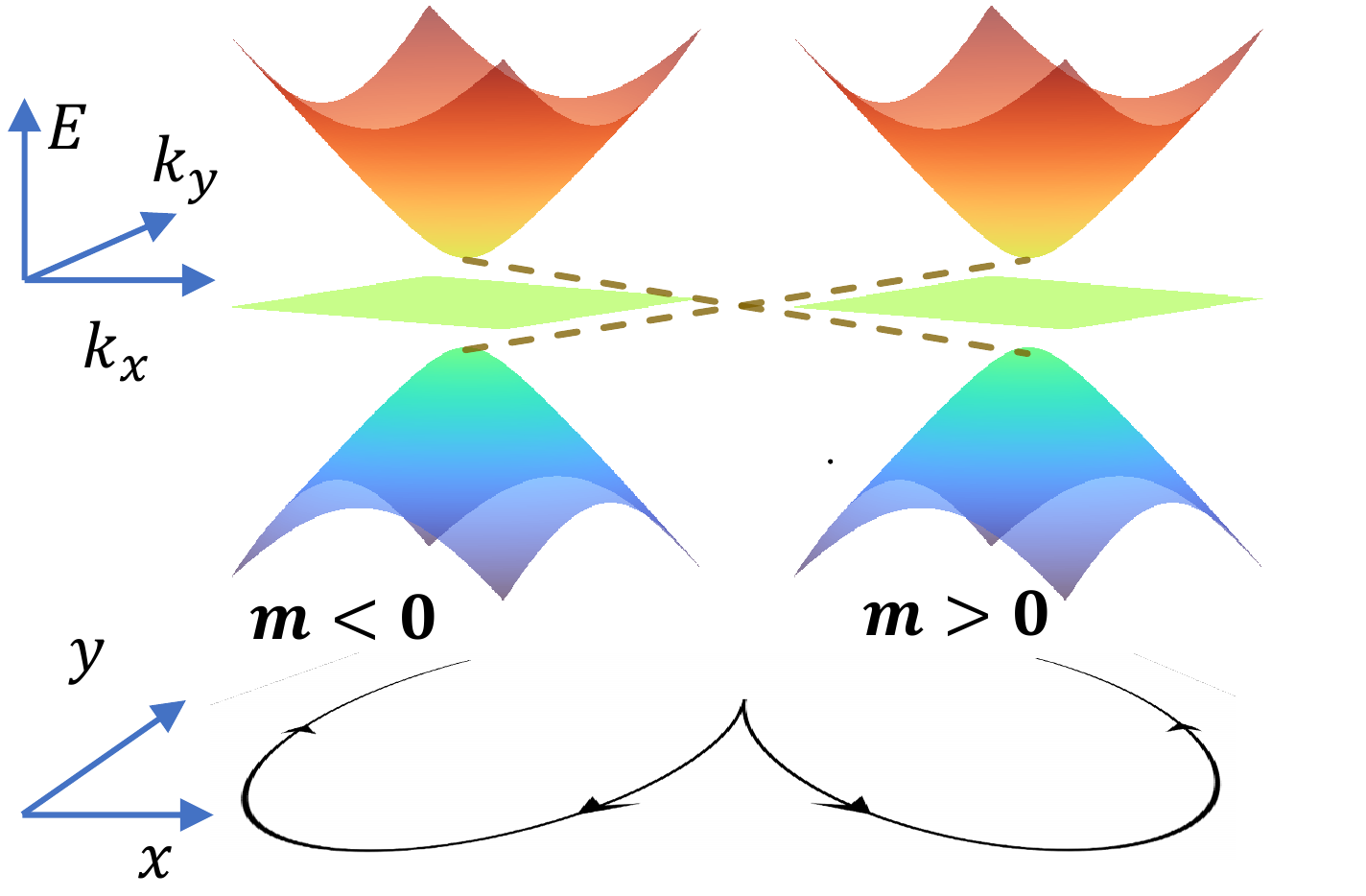}
\caption{Band inversion and corresponding reversal of ZB direction.}
\label{bi}
\end{center}
\end{figure}

Without loss of generality, we take the spin-$1$ Maxwell quasiparticle system~\cite{yqzhu} as an example to study the relationship between ZB and topological phase transition. The corresponding Bloch Hamiltonian reads $h(\mathbf{k})=\mathbf{J}\cdot \mathbf{d}(\mathbf{k}) $, where $\mathbf d=(2t_h\sin k_x,2t_h\sin k_y, 2t_h(M-\cos k_x-\cos k_y))$ and the matrices $\mathbf J$ are taken as
\begin{equation}
J_x=
\begin{pmatrix}
0&0&0\\
0&0&-i\\
0&i&0
\end{pmatrix},
J_y=
\begin{pmatrix}
0&0&i\\
0&0&0\\
-i&0&0
\end{pmatrix},
J_z=
\begin{pmatrix}
0&-i&0\\
i&0&0\\
0&0&0
\end{pmatrix}.
\end{equation}
By adjusting the parameter $M$, the system undergoes a phase transition. The band inversion occurs at the four HSPs $\mathbf k=(0,0),(0,\pi),(\pi,0),(\pi,\pi)$ and around each point the effective Hamiltonian is in the  form of Eq.~\eqref{hamJ}. For the phase transition point $M=2$, band inversion occurs at $\mathbf k=0$ and the corresponding effective Hamiltonian is
\begin{equation}
\label{effecH}
h_{(0,0)}(\mathbf{p})=v_xp_xJ_x+v_yp_yJ_y+mJ_z,
\end{equation}
where $v_x=v_y=2t_h$, $m=2t_h(M-2)$. For the intial spinor $|\Phi\rangle=(a,b,c)$ defined as
\begin{equation}
|\Phi\rangle=a|\phi_-\rangle+b|\phi_0\rangle+c|\phi_+\rangle,
\end{equation}
where $|\phi_-\rangle=\frac{1}{\sqrt{2}}(1,-i,0)^T$, $|\phi_0\rangle=(0,0,1)^T$, $|\phi_+\rangle=\frac{1}{\sqrt{2}}(1,i,0)^T$ are the eigenstates of $J_z$. Then, one can obtain
\begin{equation}
\label{xybar}
\begin{split}
\langle x_o\rangle &=\ \ \ \ \ \ \ \ \ \ \ \ \frac{v_x}{|m|}\sqrt{R^2+I^2}\sin(|m|t+\text{sgn}(m)\theta)\\
\langle y_o\rangle &=-\text{sgn}(m)\frac{v_y}{|m|}\sqrt{R^2+I^2}\cos(|m|t+\text{sgn}(m)\theta),
\end{split}
\end{equation}
where $\sin\theta=I/\sqrt{R^2+I^2}$, $\cos\theta=R/\sqrt{R^2+I^2}$, $I=\sqrt{2}\text{Im}[(a+c)b^*]$ and $R=\sqrt{2}\text{Re}[(a-c)b^*]$. We introduce the  index quantity $\nu=\text{sgn}(v_xv_ym)$  to characterize the direction of ZB, which is clockwise (counterclockwise) for $\nu=-1$ ($\nu=+1$). A change in the sign of the mass term concurs with the band inversion. Since the clockwise or counterclockwise motion of the quasiparticle is a one-to-one correspondence with band inversion, dynamical properties prove a good indicator of different topological phases (see Fig.~\ref{bi}).  In addition to that, similar band inversions can also be found at $(0,\pi)$, $(\pi,0)$ and $(\pi,\pi)$ during the phase transition. Parallel to the case of two-band system~\cite{Bernevig2013Topological}, the general expression of the topological invariant for spin-$J$ system is as follows
\begin{equation}
\label{ChJ}
\text{Ch}^J_j= -j\times\left[  \nu_{(0,0)} + \nu_{(\pi,\pi)} +\nu_{(\pi,0)} +\nu_{(0,\pi)} \right],
\end{equation}
where $j=-J,-J+1,...,J$ represents the corresponding spin indices from the lowest to the highest band, which relates the topological invariant to ZB dynamics. In Table~\ref{table1}, we list the relationship between ZB direction and the corresponding Chern number under different phases in spin-$1$ system. The mathematical proof is given in supplementary material S2.

\begin{table}[t]
	\begin{tabular}{c|c|c|c|c}
		\hline
		\hline
		$M $& $M<-2$ &$-2<M<0$ &$0<M<2$  & $M>2$ \\ \hline
		Ch & $0$ &  $2$ & $-2$ & $0$ \\ \hline\hline
		$ \nu_{(0,0)}$ & $-1$ & $-1$ & $-1$ & $+1$ \\ \hline
		$ \nu_{(\pi,\pi)}$  & $-1$ & $+1 $& $+1$ & $+1$ \\ \hline
		$ \nu_{(0,\pi)}$  & $+1$ & $+1$ & $-1$ & $-1$ \\ \hline
		$ \nu_{(\pi,0)}$  & $+1$ & $+1$ & $-1$ & $-1$ \\ \hline\hline
	\end{tabular}
	\caption{Chern number and the direction of ZB characterized by $\nu$ in different phases.}
	\label{table1}
\end{table}
For experimental purpose, it is natural to choose the Gaussian distribution as the initial state to simulate the spatial motion, namely,
\begin{equation}
\label{psi0}
|\psi(\mathbf{r},t=0)\rangle=\frac{1}{\sqrt{\pi}d}e^{-\frac{x^2+y^2}{2d^2}}\cdot |\Phi\rangle,
\end{equation}
where $d$ is the width of the wave packet and $|\Phi\rangle$ is the initial spinor. Considering the extreme situation when $d$ approaches infinity, the distribution of states in the momentum space will concentrate at $\mathbf p=0$ and one can get Eq.~\eqref{xybar}. For the case of finite width, the oscillation will decay and drift; and yet the direction of ZB stays unaffected. It is also revealed from the expression that if there is no middle component (when $b=0$), both $\emph{I}$ and $\emph{R}$ become zero and ZB disappears. This agrees well with the ZB selection rule for arbitrary spin (spin-$J$) systems (see supplementary material S1 for details). Since we consider the three-band system of spin-$1$, the middle band must exist to ensure the occurrence of ZB.

\emph{Discussion and summary.}---For non-interacting topological insulators and superconductors, an effective classification method has been developed based on the dimension and symmetry~\cite{Chiu2016}. The aforementioned three-level Bloch Hamiltonian falls into Class A, which is asymmetrical. However, the essence of the paradigm proposed in this letter lies in band inversion---a common phenomenon in the topological phase transition, which explains why our theory works well in various topological systems. To illustrate this point, let's consider two more typical classes of topological materials.

First, we consider the 2D Kane-Mele model that hosts a $\mathbb Z_2$ topological invariant in class AII~\cite{z2kanemele}. On the one hand, in the absence of Rashba spin-orbit coupling interaction, this model can be regarded as two decoupled Haldane Chern insulators with opposite Chern numbers. In this case, the phase diagram is exactly the same as that of the Haldane model~\cite{haldane}. The phase characterization of the system will be reduced to the above scenario of Class A.  On the other hand, when Rashba spin-orbit coupling is introduced, the direction of ZB stays unchanged before the phase transition occurs. The effective Hamiltonian characterizing topological phase transition and that without Rashba interaction are actually topologically equivalent, i.e., $\nu=\text{sgn}(v_xv_ym)$ remains the same as before. Therefore, ZB dynamics based on the modified Hamiltonian can still well describe the topological phase transition (see supplementary material S3 for a detailed proof and  a perturbation theory verification).

Finally, we discuss the 3D chiral topological insulators whose Hamiltonian takes the form~\cite{3dchiralti}
\begin{equation}
\begin{split}
h(\mathbf k)=&\sin k_x\lambda_4 +\sin k_y\lambda_5+\sin k_z\lambda_6\\
&+(M-\cos k_x-\cos k_y-\cos k_z)\lambda_7,
\end{split}
\end{equation}
where $\lambda_i$ is the $3\times 3$ $SU(3)$ Gell-Mann matrix. The topology of the system is protected by chiral symmetry and thus belongs to class AIII. It is worth noticing that for this three-level model, the system characterized by the 3D winding number $w$ is also mathematically equal to the so-called $\mathcal{DD}$ invariant constructed by the tensor gauge field~\cite{Palumbo2019}. Similar to the 2D system, the topological phase transition is accompanied by band inversion near the eight HSPs in the first Brillouin zone. In the vicinity of these points, the effective Hamiltonian is
\begin{equation}
h_\mathbf k=v_xp_x\lambda_4+v_yp_y\lambda_5+v_zk_z\lambda_6+m\lambda_7.
\end{equation}
To verify our theory, we calculate the trajectories of ZB before and after the topological phase transition. The results show that, unlike the 2D system, trajectories of ZB are related to the initial state. The topological invariants satisfy
\begin{equation}
w=\frac{1}{2}\sum_{i}\text{sgn}(v_xv_yv_z)_i\text{sgn}(m)_i,
\end{equation}
where the summation over $i$ corresponds to the eight HSPs. So, we need to pinpoint the signs of the parameters $v_x$, $v_y$, $v_z$ and $m$ to obtain the topological invariant. The initial state can be defined as
\begin{equation}
|\Phi\rangle=a|\phi_-\rangle+b|\phi_0\rangle+c|\phi_+\rangle,
\end{equation}
where $|\phi_0\rangle=(1,0,0)^T$, $|\phi_+\rangle=\frac{1}{\sqrt{2}}(0,1,i)^T$, $|\phi_-\rangle=\frac{1}{\sqrt{2}}(0,1,-i)^T$ are the eigenstates of $\lambda_7$. By taking the initial state of $c=0$, one can get
\begin{equation}\label{xyz1}
\begin{split}
\langle x_o(t)\rangle&=-\sqrt{2}\frac{v_x}{m}R_2\cos(mt+\theta_2)\\
\langle y_o(t)\rangle&=-\sqrt{2}\frac{v_y}{m}R_2\sin(mt+\theta_2)\\
\langle z_o(t)\rangle&=0,
\end{split}
\end{equation}
where $R_2e^{i\theta_2}\equiv ab^*$. In this case, one can extract the sign of $v_xv_ym$ by observing the ZB dynamics, which can determine the topological invariant up to an overall sign (see S4 in spplementary material).  Next, we take the initial state of $b=0$ and the ZB trajectory turns into
\begin{equation}\label{xyz2}
\begin{split}
\langle x_o(t)\rangle&=\langle y_o(t)\rangle=0\\
\langle z_o(t)\rangle&=-\frac{v_z}{m}R_3\cos(2mt+\theta_3),
\end{split}
\end{equation}
where $R_3e^{i\theta_3}\equiv ac^*$. By combining Eq.~\eqref{xyz1} and \eqref{xyz2}, corresponding topological invariants can be obtained.

In summary, the link between ZB and topological invariants in a general spin-$J$ system has been established, and the core of this link is that the band inversion during topological phase transition will reverse the direction of ZB. Despite the many variations in topology, band inversion constitutes the only defining feature of topological insulators and topological superconductors. So far, ZB has been proposed or realized in lots of table-top setups such as cold atomic gases~\cite{njp15073011,pra88021604,hasan2022anisotropic,PhysRevLett.100.153002}, photonic crystal~\cite{prl100113903}, optical waveguide array~\cite{prl105143902}, twisted bilayer systems~\cite{PhysRevLett.127.106801}, superconducting circuit~\cite{pra99042308}, circuit-QED setups~\cite{Pedernales_2013,PRX2021007} and so on. Supported by the various experimental platforms, the theory in this letter will find wide application in topological materials. 

Representative examples are given to illustrate the universality of the theory, inclusive of Class A (Maxwell quasiparticle), Class AII (Kane-Mele model) and Class AIII (3D Chiral topological insulators), which all show validity of this theory. In particular, as for the $\mathbb Z_2$ invariant, there is currently no effective detection approach in experiments. Here, our research of Kane-Mele model suggests that $\mathbb Z_2$ invariants can be measured through ZB dynamics. Moreover, this dynamical method also provide a new way to detect the 3D winding number and $\mathcal{DD}$ invariant from the macro perspective. Furthermore, considering the current progress in the quantum simulation experiments~\cite{XTan2018,WJi2020,TXin2020,XTan2021,MChen2021}, this method can also be applied to the corresponding topological semimetallic phases, such as 3D spin-1 Maxwell (spin-$J$) semimetals~\cite{yqzhu}, Weyl semimetal~\cite{PhysRevA.94.043617}, 3D Dirac semimetals, 4D tensor semimetals~\cite{Palumbo2018,YQZhu2020}, etc. 

In a word, this work not only develops a general theory of spin-$J$ systems, but also provides a new scheme to measure topological invariants from the perspective of quasiparticle dynamics.

\emph{Acknowledgements.}---X.S. acknowledges the support by NSFC (Grant No. 12104430). Z.L. acknowledges the support by NSFC (Grants No. 11704132), NSAF (Grant No. U1830111), the Natural Science Foundation of Guangdong Province (No. 2021A1515012350), and the KPST of Guangzhou (Grant No. 201804020055).

\bibliographystyle{apsrev4-1}
\bibliography{ref.bib}

%
%
%
%
\end{document}